\documentclass[aps,superscriptaddress]{revtex4}
\usepackage{natbib}
\usepackage{color}
\usepackage{amssymb}
\usepackage{amsmath, amsthm}
\usepackage{epsfig}
\usepackage{epstopdf}
\newcommand{\ket}[1]{\vert#1\rangle}

\usepackage{graphicx}
\usepackage{bm}

\begin{document}

\title{Resilience of hybrid optical angular momentum qubits to turbulence}

%%%%%%%%%%%%%%%%%%%%%%%%%%%%%%%%%

%---------------------------------------
\author{Osvaldo Jim\'enez Far{\'\i}as}
\email{ofarias@if.ufrj.br}
\affiliation{Instituto de F\'{\i}sica, Universidade Federal do Rio de
Janeiro, Caixa Postal 68528, Rio de Janeiro, RJ 21941-972, Brazil}
%---------------------------------------

\author{Vincenzo D'Ambrosio}
\affiliation{Dipartimento di Fisica, Sapienza Universit\`a di Roma, piazzale Aldo Moro 5, 00185 Roma, Italy}

\author{Caterina Taballione}
\affiliation{Dipartimento di Fisica, Sapienza Universit\`a di Roma, piazzale Aldo Moro 5, 00185 Roma, Italy}

\author{Fabrizio Bisesto}
\affiliation{Dipartimento di Fisica, Sapienza Universit\`a di Roma, piazzale Aldo Moro 5, 00185 Roma, Italy}
\author{Sergei Slussarenko}
\altaffiliation[Current address: ]{Centre for Quantum Dynamics, Griffith University, 170 Kessels Road QLD 4111, Australia}
\affiliation{Dipartimento di Fisica, Universit\`a di Napoli Federico II, Compl. Univ. di Monte S. Angelo, 80126 Napoli, Italy}
\author{Leandro Aolita}
\affiliation{Dahlem Center for Complex Quantum Systems, Freie Universit\"at Berlin, Berlin, Germany}

\author{Lorenzo Marrucci}
\affiliation{Dipartimento di Fisica, Universit\`a di Napoli Federico II, Compl. Univ. di Monte S. Angelo, 80126 Napoli, Italy}
\affiliation{CNR-SPIN, Complesso Universitario di Monte S. Angelo, 80126 Napoli, Italy}

\author{Stephen P. Walborn}
\affiliation{Instituto de F\'{\i}sica, Universidade Federal do Rio de
Janeiro, Caixa Postal 68528, Rio de Janeiro, RJ 21941-972, Brazil}
%---------------------------------------
\author{Fabio Sciarrino}
%\email{fabio.sciarrino@uniroma1.it}
%\homepage{http://quantumoptics.phys.uniroma1.it}
\affiliation{Dipartimento di Fisica, Sapienza Universit\`a di Roma, piazzale Aldo Moro 5, 00185 Roma, Italy}
\affiliation{Istituto Nazionale di Ottica (INO-CNR), Largo E. Fermi 6, I-50125 Firenze, Italy}

%---------------------------------------

\date{\today}

\begin{abstract}
\textbf{Recent schemes to encode quantum information into the total angular momentum of light, defining rotation-invariant hybrid qubits composed of the polarization and orbital angular momentum degrees of freedom, present interesting applications for quantum information technology.  However, there remains the question as to how detrimental effects such as random spatial perturbations affect these encodings. Here, we demonstrate that alignment-free quantum communication through a turbulent channel based on hybrid qubits can be achieved with unit transmission fidelity. In our experiment, alignment-free qubits are produced with $q$-plates and sent through a homemade turbulence chamber. The decoding procedure, also realized with $q$-plates, relies on both degrees of freedom and renders an intrinsic error-filtering mechanism that maps errors into losses.}
\end{abstract}

\pacs{}

\maketitle

%\section{Introduction}
Successful transmission of information from one location to another depends on the physical system carrying the information and how robust it is under the unavoidable influence of the environment.  Among all the physical candidates for carriers of quantum information, light represents the most reliable one for long-distance communication, due to its low interaction with the environment and its high speed. The most popular degree of freedom of light exploited in quantum technologies is the polarization of single photons. However, polarization encoding requires a shared reference frame between sender and receiver \cite{peres01}, which represents a limitation, specially if one considers scenarios where the relative reference frame changes, as for example in satellite communication.  This obstacle can be overcome with ``hybrid" rotational invariant single photons states where qubits are encoded into both the polarization and orbital angular momentum (OAM) degrees of freedom \cite{aolita07,souza08,dambrosio12}.  Recent progress on the control of the orbital OAM of light has proven it  also as a good candidate for storage, transmission, and manipulation of quantum information with promising applications  \cite{molina-terriza07,nagali09,franke-arnold08,dambrosio13,jha11,fickler12}. Meanwhile, quantum technologies with hybrid systems develop rapidly unveiling new applications in metrology \cite{ambrosio13}, and more recently in quantum key distribution through free space \cite{vallone14}.     However,  spatial properties  of light such as OAM are severely affected by spatial perturbations such as the optical turbulence present in the atmosphere \cite{gibson04,Paterson05,smith06,walborn08b,pors11,roux11,malik12, Ibrahim13,pereira13,ren13}. OAM optical systems encoding classical information have recently started to be tested in real scenarios over distances of the order of kilometers with positive results and perpectives for the transmission of quantum information \citep{krenn14}.

Since hybrid qubits exploit OAM properties, they are also susceptible to atmospheric turbulence.
This raises the question of whether the use of this additional degree of freedom does indeed provide a practical advantage.  Here we show that, in the weak turbulence regime,  unity transmission fidelity of rotational invariant hybrid qubits can in principle be achieved.  This robustness is due to both the logical encoding and the decoding procedure, which uses a ``q-plate" device \cite{marrucci06,nagali09} and a projection onto the fundamental gaussian mode with a single-mode fiber. A preliminary study was reported in Ref. \cite{dambrosio12}, where it was predicted that spatial perturbations mirror symmetric in OAM space should result only in losses, rather than qubit state errors. One can gain intuition of this intrinsic error correction procedure as follows: Hybrid qubits employ polarization and OAM eigenstates $\ket{l}$, $\ket{-l}$.  The wavefunctions of these OAM states are equivalent up to a mirror reflection, so a spatial perturbation with this symmetry affects both components of OAM equally.  In the decoding procedure, the projection onto a zero-order OAM mode transfers the encoded information to the polarization degree of freedom in a robust fashion: The effects of the OAM perturbation would factor as they affected both OAM states equally.   

Nevertheless, a turbulent atmosphere perturbs the spatial structure of a beam by tilting, skewing, displacing or a combination of all of these in a stochastic way. In general, such effects do not respect any symmetry.  However, in the weak turbulence regime the perturbation appears as a multiplicative spatially-dependent phase term \cite{Paterson05}. This is enough to guarantee mirror symmetry, so that our decoding method  selects only the homogeneous contribution OAM terms, resulting in high transmission fidelity and mapping all the inhomogeneous contributions into losses.

Here, we quantify these losses and perform an experimental demonstration of the fidelity resilience under atmospheric turbulence produced with a turbulence chamber which is similar to those used for simulating conditions of observation in astrophysics \cite{keskin06}.  These results show that the hybrid encoding is a potential resource for practical long-distance quantum communication without a shared reference frame, as an Earth-to-satellite communication protocol would require.

\section*{Results}
 Let us start by considering the effect of turbulence on the unperturbed OAM eigenstates of a monochromatic paraxial field in the cylindrical basis: 
\begin{equation}
\varphi_{l}(r,\theta)= \langle r,\theta| l\rangle=\frac{1}{\sqrt{2\pi}}\langle r|l\rangle e^{il \theta},
\end{equation} 
where $r$ is the radial coordinate, $\theta$ is the azimuthal angle and $l$ is the orbital angular momentum quantum number.
For simplicity we will omit the explicit dependence of the wave function on the propagation distance $z$. We consider turbulence in the so-called weak regime, where the random fluctuations of the index of refraction cause only phase aberrations in the transverse profile \cite{Paterson05}. This means that an initial OAM eigenstate transforms as
\begin{eqnarray}
\varphi_{l}(r,\theta) \rightarrow \psi_{l}(r,\theta)=\varphi_{l}(r,\theta)e^{i\phi(r,\theta)},\label{perturbed}
\end{eqnarray}
where $\phi(r,\theta)$ is the random variable associated to the phase perturbation. In general, $ \psi_{l}(r,\theta)$ is a superposition of several OAM eigenstates.  Kolmogorov's  theory of turbulence \cite{andrews} describes the stochastic properties of this perturbation by means of the so called coherence function
\begin{equation}
\langle e^{i\phi(r,\theta)-i\phi(r,\theta')}\rangle_{ens}=e^{-6.88\times 2^{\frac{2}{3}}(\frac{r}{r_0})^{\frac{5}{3}}|\sin(\frac{\theta-\theta'}{2})|^{\frac{5}{3}} }, \label{ensemble}
\end{equation} 
where the subscript ``$ens$'' denotes ensamble average. The strength of turbulence in this regime is measured by the ratio between the beam waist $w$ and the Fried parameter ${r_0}$,   given by \cite{fried66}
\begin{equation}
r_0=0.185 \left(\frac{\lambda^2}{C_n^2 z} \right)^{3/5},
\end{equation}
where $\lambda$ is the wavelength and $C_n$ is the refractive index structure constant. It
can be associated with the coherence length of the phase perturbations: the smaller the value of $r_0$ compared to $w$,  the stronger the effects of turbulence.  
\par

Rotationally invariant photonic qubits are composed of polarization and $\ell=\pm 1$ OAM eigenstates. However in some cases one might be interested in transmitting higher order hybrid qubits made of polarization and opposite OAM values with $|l|>1$, as was done in \cite{ambrosio13}. The calculations we present here, concerning only the OAM part, can be applied in the future to higher order hybrid qubits. For this reason we consider a two-dimensional OAM subspace spanned by states $\ket{-l}$ and $\ket{l}$.  Moreover, we assume that the detection is such that only these states are registered.   That is, any perturbation that takes the OAM outside of the subspace results in a null event, or a loss.  This of course depends on the detection procedure, and can be true in the case of pure OAM \cite{Ibrahim13} or hybrid qubits \cite{dambrosio12}.  

Thus, in order to know the  properties of the evolved state after turbulence, we have to calculate matrix elements of the form $\langle l_1|\psi_{I}\rangle\langle\psi_{II}|l_2\rangle$ where the OAM quantum numbers $l_{1(2)}$ and $l_{I,(II)}$ take values $\pm l$. The basis elements labeled by arabic numerals $\ket{l_1}$ and $\ket{l_2}$ can be interpreted as the states at the measurement stage. Roman numerals denote the initially prepared states $\ket{l_{I}}$ and $\ket{l_{II}}$ which after turbulence become $\ket{\psi_I}$ and $\ket{\psi_{II}}$ according to equation (\ref{perturbed}). Following the results in  \cite{Paterson05}, we proceed to calculate the ensemble average where the non-zero elements can be written as 
\begin{equation}
C_{\Delta l} =\langle \langle l_1|\psi_{I}\rangle\langle\psi_{II}|l_2\rangle\rangle_{ens}= \int|R(r,z)|{^2} \Theta_{\Delta l} (r,\theta,\theta') rdr,
 \label{ces}
\end{equation}
where $R(r,z)$ is the radial distribution of the unperturbed wave front and the function $\Theta_{\Delta l }$ is the circular harmonic transform of the rotational coherence function 
\begin{equation}
\Theta_{\Delta l} (r,\theta,\theta')=\int e^{-i \Delta  l{(\theta-\theta')}} \langle e^{i\phi(r,\theta)-i\phi(r,\theta')}\rangle_{ens} d\theta d\theta',  \label{teta}
\end{equation}
with $l_I -l_1-(l_{II}-l_2)=\Delta l$. This quantity must be $\Delta l=0,2l$ for preparation and detection in the subspace spanned by OAM values $\pm l$.
\begin{figure}
\begin{center}
\includegraphics[width=15cm]{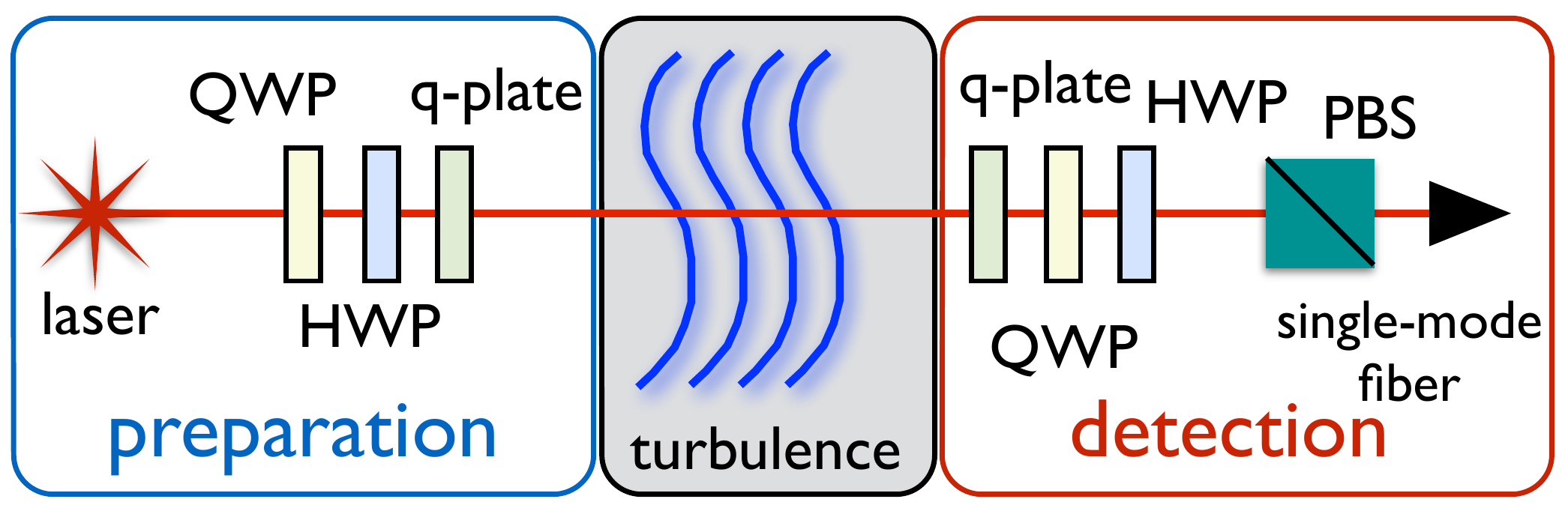}
 \caption{\textbf{Basic qubit encoding and decoding procedures considered for hybrid polarization/OAM qubits.}  A linear-polarized single-mode gaussian beam is transformed into a hybrid qubit using a quarter-wave-plate (QWP), half-wave plate (HWP) and $q$-plate device.  The decoding procedure works in reverse, and the output beam is sent through a polarizing beam splitter (PBS) coupled into a single-mode fiber.}
\label{fig:basic}
\end{center}
\end{figure} 
 
From equations (\ref{ces}) and (\ref{teta})  one can see that the non-zero elements are
\begin{equation}
C_0 =\langle l|\psi_l\rangle\langle\psi_{-l}|-l\rangle=|\langle\psi_l|l\rangle|^2=|\langle\psi_{-l}|-l\rangle|^2 \label{c0}
\end{equation}
and 
\begin{equation}
C_{2l}=|\langle\psi_{-l}|l\rangle|^2=|\langle\psi_l|-l\rangle|^2. \label{c2} 
\end{equation}
We note that Eq. \eqref{c0} satisfies the mirror symmetry condition found in \cite{dambrosio12} for resilience against spatial perturbations. Coefficients $C_0$ and $C_{2l}$ depend on $w/r_0$, but we will omit this dependence for notational simplicity.

\textbf{Hybrid Qubits.} In order to analyze the effect of turbulence on the hybrid qubits, we consider a quantum communication scenario without a shared reference frame as in Fig. \ref{fig:basic}. Logical qubits are encoded into the subspace of the polarization/OAM degrees of freedom spanned by the vectors:  
\begin{equation}
\begin{array}{l}
\ket{0}_{h} =\ket{R,l} \\
\ket{1}_{h} =\ket{L,-l},
\end{array}
\label{hybrid}
\end{equation}
where $h$ stands for ``hybrid" and  $R(L)$ stands for right (left) circular polarization. One can efficiently produce arbitrary superpositions of states in equations \eqref{hybrid} by first encoding qubits in the polarization via waveplates and then by transferring them to the hybrid space by means of a $q$-plate, an inhomogeneous birefringent device that couples the  polarization and  OAM \cite{marrucci06,nagali09}. The $q$-plate acts indeed as a universal encoder/decoder, mapping back and forth between the polarization qubit $\ket{\Psi}=\alpha\ket{R}+\beta\ket{L}$ and the hybrid qubit  $\ket{\Psi}_{h}=\alpha \ket{0}_h + \beta \ket{1}_h$ \cite{dambrosio12}.  

When $l=1$, the hybrid states $\ket{0}_h$ and $\ket{1}_h$ have null total angular momentum, and are invariant to rotations around the propagation axis \cite{aolita07,dambrosio12}.   
Hybrid qubits are then sent through a turbulent environment. 
For the OAM component  in equations (\ref{hybrid}), after turbulence we have 
\begin{eqnarray}
\ket{\pm l}_{oam} \rightarrow \ket{\psi_{\pm l}}_{oam},
\end{eqnarray}
where in general $\langle \psi_i \ket{\psi_j}_{oam}\neq 0$ even for $i\neq j$, which means that orthogonality in the OAM basis is not preserved by turbulence.
\par
In turn, the effects of turbulence on the polarization degrees of freedom are negligible. So, for hybrid states of light in Eq.(\ref{hybrid}) after transmission we have
\begin{equation}
\begin{array}{l}
\ket{0}_{h}  \rightarrow \ket{0^T}_{h}  = \ket{R}\ket{\psi_{-l}} \\ 
\ket{1}_{h}  \rightarrow \ket{1^T}_{h}  =\ket{L}\ket{\psi_{l}},
\end{array}
\end{equation}
where ``$T$" stands for turbulence.  Because the polarization is immune to turbulence, orthogonality between logical states is preserved even after transmission through turbulence.  %after transmission through turbulence. 

 \begin{figure}
\includegraphics[width=17cm]{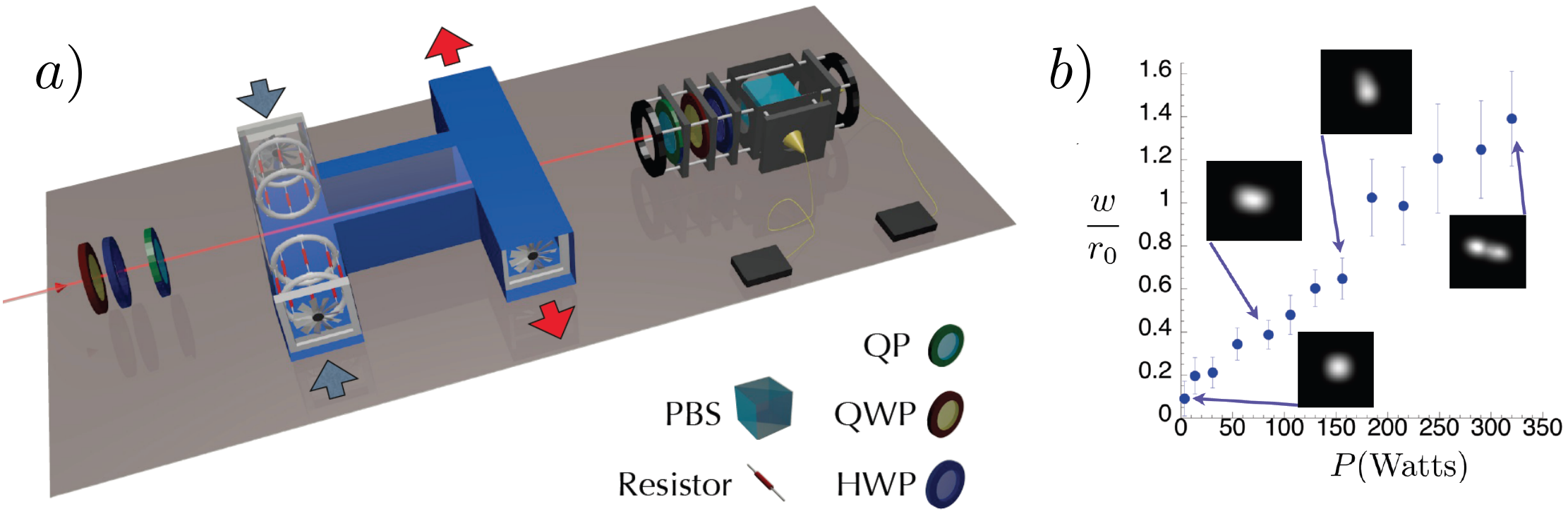}
\caption{ \textbf{ Experimental setup and calibration of the turbulence chamber}  a) Experimental setup.  A laser beam is sent through an encoding stage, consisting of a QWP, HWP and q-plate, used to encode a hybrid qubit.  The hybrid qubit is then sent through a homemade turbulence chamber, mixing hot and cold air (see text).  The decoding stage consists of a q-plate and a polarization analysis stage. Light is coupled into single-mode fibers and sent to a photodiode.  We also use a CCD camera (not shown) to characterize the effects of turbulence.  The decoding stage is mounted on a rotation stage, to simulate misalignment.  b) Characterization of turbulence induced by the chamber given by the ratio of the beam waist $w$ to the Fried parameter $r_0$  as a function of the power applied to the resistors in the chamber.  The insets show snapshots of the beam profile taken with a CCD camera with 1ms exposure time.}\label{fig:setup}
\end{figure}

Finally, the detection stage consists of a second $q$-plate, wave plates for polarization analysis and a single mode fiber. The results (\ref{c0}) and (\ref{c2}) of the last section can be put together to conclude that after the $q$-plate, the state transforms as \cite{dambrosio12}: 
  \begin{align}
  \label{eq:decode}
\alpha \ket{0^T}_h + \beta \ket{1^T}_h  \xrightarrow{q-plate} &\sqrt{P_h}(\alpha \ket{R}+\beta\ket{L})\ket{l=0} \\ \nonumber 
& + \mathrm{higher\,order}\, l \, \mathrm{terms},
\end{align}
where $P_h$ is the probability of the perturbed state to remain inside the logical subspace. By post-selecting the fundamental $l=0$ spatial mode, the quantum state can be perfectly recovered and encoded back into the polarization space. Thus, unity fidelity is in principle possible for the hybrid encoding and any amount of weak turbulence. This post-selection can be made in a non-destructive way, by simply sending the photon through a single-mode optical fiber.  Then, the polarization state can be detected using waveplates and a polarizing beam splitter, as shown in Fig. \ref{fig:basic}, or used for further quantum-information processing.

$P_{h}$ constitutes the probability of success of the protocol, and is given by $P_{h}=C_0$. Explicitly, using equations \eqref{ensemble}, \eqref{ces} and \eqref{teta}, we have
\begin{equation}
P_h= \iint|R(r,z)|{^2} e^{-6.88\times 2^{\frac{2}{3}}(\frac{r}{r_0})^{\frac{5}{3}}|\sin(\theta-\theta^\prime)|^{\frac{5}{3}} } r dr d\theta d\theta' .
 \label{Ph}
\end{equation}

\textbf{Experimental Implementation.} The experimental setup used to demonstrate the resilience of hybrid qubits is sketched in Fig. \ref{fig:setup}.  A linear polarized laser beam with wavelength $\lambda=795$ nm propagates first through a preparation stage consisting of a quarter wave plate (QWP) a half wave plate (HWP) and a $q$-plate where arbitrary superpositions of hybrid states can be prepared. A turbulence machine (described in the Methods section) produces real air turbulence between the preparation and detection stages. In Fig. \ref{fig:setup} b), we observe the relation between power at the resistors inside the turbulence machine  and the Fried parameter characterizing the turbulence strength. Progressive degradation of a gaussian beam is illustrated by 1 ms snapshots for four different Fried parameters. After transmission through turbulence, the information is decoded by a $q$-plate, followed by QWP, HWP and PBS and the light is coupled to a single-mode fiber and detected.
\begin{figure}
\begin{center}
\includegraphics[width=17cm]{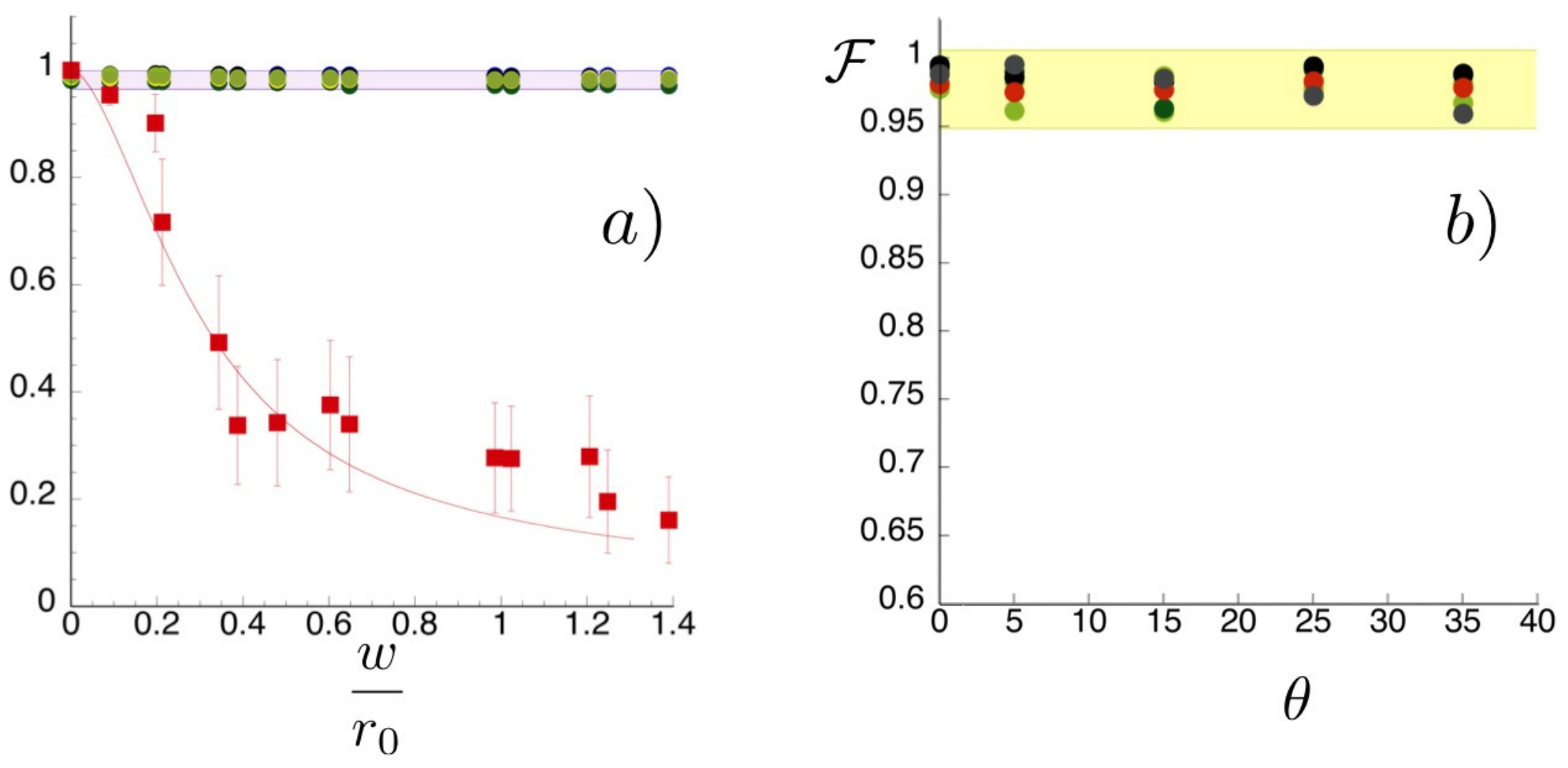}
\caption{ \label{fig:fide} \textbf{Experimental results for fidelity and losses} a) Success probability  (red squares) and fidelity of hybrid states (circles) as a function of the  turbulence strength $w/r_0$. For the fidelities, each color corresponds to an eigenstate in mutually unbiased bases; the grey area encloses the region of dispersion of the experimental points around the mean value 0.985. We can observe the resilience to random phase perturbations. b)The rotational invariance of hybrid qubits is preserved in the weak turbulence regime. Fidelities do not depend on the angle $\theta$ between sender and receiver's frames, for $\frac{w}{r_0}=0.6$. The yellow area is the region of dispersion of the experimental points.}
\end{center}
\end{figure}

%:
For 14 different strengths of turbulence, we prepare the 6 states of the set of mutually unbiased bases of dimension 2. Fidelity with respect to the initial states is measured at the detection stage obtaining a total of 84 points distributed around a mean value of $\mathcal{F}_h=0.985\pm 0.006$. The experimental results are shown in Fig. \ref{fig:fide}: Fidelity of transmission for the initial states $\ket{0}_h$ (orange), $\ket{1}_h$ (yellow), $\ket{\pm}_h=(\ket{0}_h\pm \ket{1}_h)/\sqrt{2}$ (blue, green), $\ket{R}_h=(\ket{0}_h-i\ket{1}_h)/\sqrt{2}$ (black), and $\ket{L}_h=(\ket{0}_h+i\ket{1}_h)/\sqrt{2}$ (red). The effective coupling efficiency decays as the turbulence increases giving a measure of the probability of post selection $P_h$, shown by the red squares in Fig. \ref{fig:fide}. The red solid curve is a plot of $P_h$ given in equation. \eqref{Ph}, and showing good agreement between experiment and theory. 

To show that the resistance to turbulence is valid in conjunction with the alignment-free character of the hybrid qubits, we repeat the experiment for the same set of states but this time we rotate the detection kit along the optical axis at several angles. The strength of turbulence was locked at $\frac{w}{r_0}=0.6$. Experimental results are shown in the inset of Fig. \ref{fig:fide}. Again we can conclude the resilience of the qubits  by observing a mean fidelity of $\mathcal{F}_h=0.98\pm0.01$ taken over 30 experimental points, for several different alignment angles. 
 
\section*{Discussion}

 We have shown both theoretically and experimentally that rotational-invariant hybrid photonic qubits, encoded in the polarization and orbital angular momentum degrees of freedom, are resistant to atmospheric turbulence.  The robustness is due to both the logical encoding and the decoding procedure, realized with a $q$-plate device, which provides an intrinsic error-filtering mechanism.  Our results show that quantum communication without a shared reference frame schemes using hybrid qubits are viable over long distances in free-space.    

We emphasize that our theoretical predictions are based on KolmogorovÕs theory of turbulence. Our experimental data show excellent agreement with these predictions. The same theory was used in other experiments where thick turbulence was simulated by phase screens \cite{rodenburg14}. Besides, recent experiments in real life situations, as transmission of OAM states of light over a 3 km distance, demonstrate that, even in this cases, Kolmogorov's theory is still valid \cite{krenn14}. This indicates that there are no fundamental obstacles to the application of the present proof-of-principle results to this type of real-life situations.  On the other hand, it is known from astronomical observations that stratospheric turbulence deviates from the predictions of the Kolmogorov model \cite{mikhal97}. Thus, non-Kolmogorov effects will necessarily have to be taken into account for Earth- to-satellite quantum communication links. We believe our results constitute a basis for future research in that direction.

\section*{Methods}

In order to produce a turbulent medium we construct a machine that causes random fluctuations of the index of refraction in the air using a similar design to that used in Refs. \cite{keskin06,pors11}.  The light beam is sent through an air chamber that is 20 cm long, with total volume 220 cm$^3$.  Two cylindrical arrays of resistors with a total electrical resistance of 15 Ohms are driven at a maximum power of 290W, heating the air inside the chamber and producing a temperature difference of up to $\Delta T=230^\circ$C with respect to the room temperature (about $18^\circ$C).  Two fans behind the resistors produce the dynamical turbulence by mixing the air at the room temperature with the heated air, resulting in fluctuations of the index of refraction.     
 
In order to calibrate our machine we place a CCD camera at the waist $w$ of a Gaussian beam after the turbulence chamber and observe the average broadening of the Gaussian profile after propagation in the turbulence chamber, as  compared to the unperturbed beam. For a given value of the power applied to the resistors, we record a 20 s video using the CCD camera.  Summing over all frames of the video, we measure the width of the Gaussian distributions with and without turbulence.  Following Refs. \cite{fante75,pors11}, the ratio $w/r_0$ can be obtained via the equation
\begin{equation}
\frac{w}{r_0}=\frac{1}{3}\sqrt{\left(\frac{w_{t}}{w}\right)^2 -1}, 
\end{equation}
where $w_t$ is the width of the beam after turbulence.   This allows us to measure values from 0 to a maximum value of $ \frac{w}{r_0}=1.4 \pm 0.2$, corresponding to a wide range of the turbulence regime.

%\bibliographystyle{naturemag}
%%%\bibliography{/Users/spwalborn/Dropbox/Master_Bibtex}
%\bibliography{Master_Bibtex}

\section*{Acknowledgments}

This work was supported by the FET-Open Program, within the 7th Framework Programme of the European Commission under Grant No. 255914, the EU Marie Curie Grant IEF No. 299141, PHORBITECH, FIRB-Futuro in Ricerca (HYTEQ), EU project  QWAD (Quantum Waveguides Applications and Development) and the brazilian agencies CNPq, FAPERJ and the INCT-Informa\c{c}\~ao Qu\^antica. We would like to thank C. H. Monken and S. Giacomini for helpful discussions about the turbulence machine.

\section*{Author Contribution}
O.J.F., V.D., L.A., L.M., S.P.W.,  and F.S., conceived the experiment. O.J.F. and S.P.W. designed and built the first turbulence machine prototype. O.J.F., V.D., C.T., F.B., S.P.W., and F.S. improved and optimized the turbulence machine. O.J.F., V.D., C.T., F.B., and F.S. carried out the experiment and analyzed the data. O.J.F., V.D., C.T., F.B., S.S., L.A., L.M., S.P.W., and  F.S. discussed the data and wrote the manuscript.

\section*{Additional Information}

\textbf{Competing finantial interests:} The authors declare no competing finantial interests.\\

\end{document}